\documentclass[aps,prd,twocolumn,superscriptaddress,nofootinbib]{revtex4}
\pdfoutput=1
\usepackage{amsmath}
\usepackage{siunitx}

\usepackage{graphicx}
\usepackage[section]{placeins}
\usepackage[usenames]{xcolor}

\usepackage{hyperref}

\newcommand{\Su}{{\(S_\uparrow\)}}
\newcommand{\So}{{\(S_0\)}}
\newcommand{\Sd}{{\(S_\downarrow\)}}

\begin{document}

\title{Gravitational Waves from Disks Around Spinning Black Holes: Simulations in Full General Relativity}

\author{Erik Wessel}
\affiliation{Department of Physics, University of Arizona, Tucson, AZ}

\author{Vasileios Paschalidis}
\affiliation{Departments of Astronomy and Physics, University of Arizona, Tucson, AZ}

\author{Antonios Tsokaros}
\author{Milton Ruiz}
\affiliation{Department of Physics, University of Illinois, Urbana-Champaign, IL}
\author{Stuart L. Shapiro}
\affiliation{Departments of Physics and Astronomy, University of Illinois, Urbana-Champaign, IL}

\date{\today}

\begin{abstract}
We present fully general-relativistic numerical evolutions of
self-gravitating tori around spinning black holes with dimensionless
spin \(a/M = 0.7\) parallel or anti-parallel to the disk angular
momentum. The initial disks are unstable to the hydrodynamic
Papaloizou-Pringle Instability which causes them to grow persistent
orbiting matter clumps. The effect of black hole spin on the growth
and saturation of the instability is assessed. We find that the
instability behaves similarly to prior simulations with non-spinning
black holes, with a shift in frequency due to spin-induced changes in
disk orbital period. Copious gravitational waves are generated by
these systems, and we analyze their detectability by current and
future gravitational wave observatories for large range of masses. We find that systems of \(10 M_\odot\) - relevant for black
hole-neutron star mergers - are detectable by Cosmic Explorer out to
\(\sim300\) Mpc, while DECIGO (LISA) will be able to detect systems of
\(1000 M_\odot\) ($10^5M_\odot$) - relevant for disks forming in
collapsing supermassive stars - out to cosmological redshift of
\(z\sim5\) ($z\sim 1$). Computing the accretion rate of these systems
we find that these systems may also be promising sources of coincident
electromagnetic signals.
\end{abstract}


\maketitle

\section{Introduction\label{section:intro}}

To realize the full potential of multimessenger astronomy, it is
necessary to model a wide range of gravitational wave (GW)
sources. With planned sensitivity upgrades to Advanced LIGO, KAGRA,
and Virgo \cite{Collaboration2013} as well as the construction of
future more sensitive detectors such as Cosmic Explorer
\cite{Reitze2019} and the Einstein Telescope \cite{Punturo2010}, a
large volume of space will be opened up to GW astronomy in the coming
years.  Space-based missions such as DECIGO \cite{Sato2017} and LISA
\cite{AmaroSeoane2017} will open up lower frequencies extending down
to \(0.1 \mathrm{Hz}\) and \(10^{-4} \mathrm{Hz}\), respectively,
which are inaccessible from the ground and will enable the potential
detection of entirely new types of sources. With such a large volume
of space and such a wide range of frequencies on the verge of being
observed, we should be prepared to detect the unexpected, especially
once detectors of sufficient sensitivity are brought on-line. It is
therefore prudent to investigate sources that have not yet received
significant consideration.

While much modeling has been done for compact binary coalescences (for
reviews, see
\cite{Lehner2014,Shibata2011,Faber2012,Paschalidis2017,Baiotti2017,Paschalidis2017a,Duez2018,Ciolfi2020}),
comparatively little work has been done exploring the multi-messenger
signatures of black holes (BHs) surrounded by massive accretion disks.
These systems can arise in various astrophysical environments. For
example, disks with rest-masses $\gtrsim 10\%$ of the BH Christodoulou
mass can form following black hole-neutron star mergers with rapidly
spinning black holes~\cite{Lovelace2013} or in the collapse of
supermassive
stars~\cite{Shibata2002,Shapiro2002,Shapiro2004,Shibata2016a,Sun2017,Uchida2017,Sun:2018gcl},
and possibly also in
collapsars~\cite{MacFadyen:1998vz,MacFadyen1999,MacFadyen2001,Heger2002,Heger2003}.
Binary neutron star systems with large mass asymmetry can also produce
massive disks \cite{Rezzolla_2010}. Accretion disks onto black holes
can be hosts of a wide range of dynamical instabilities that can
produce a time-varying quadrupole moment, making them promising GW
candidates. In addition, such systems can generate bright
electromagnetic signals, and hence they are true multimessenger
sources.

A prime example of a dynamical disk instability that develops a
time-varying quadrupole moment is the so-called Papaloizou-Pringle
Instability (PPI).  The PPI is a hydrodynamic instability that grows
in fluid tori orbiting in a central potential \cite{PP84}. It results
in the growth of non-axisymmetric modes in the rest-mass density
$\rho_0$ of the form
\begin{equation}\label{eq:mode_form}
\rho_0 \propto e^{i(m \phi - \sigma t)},
\end{equation}
where \(m\) is a positive mode number, and \(\sigma\) has a real
component that causes pattern rotation and an imaginary component that
causes growth. The growth rate is on the order of the orbital
timescale of the disk, with \(m=1\) being the dominant mode for thick
tori \cite{Kojima86a}. For low-\(m\) modes in tori of finite extent,
exponential growth can be thought of as resulting from the exchange of
a conserved quantity between wave-like disturbances on the disk's
inner edge that propagate opposite to the flow, and wave-like
disturbances on the disk's outer edge that propagate with the flow
\cite{Blaes1985,Blaes1986,Goldreich1986,Narayan1987,Goodman1988,Christodoulou1992}. 
The existence of unstable PPI modes depends on the
profile of a disk's specific angular momentum, \(j\). In a Newtonian
context, \(j\) is defined as
\begin{equation}
j \equiv r v_\phi,
\end{equation}
where \(r\) is the cylindrical coordinate and \(v_\phi\) is the fluid velocity in 
the \(\phi\)-direction. In \cite{PP85} it was shown that disks are susceptible 
to the instability when they have a sufficiently shallow \(j\) profile:
\begin{equation}
j \propto r^{2-q},\; \sqrt{3} < q \leq 2,
\label{eq:newj}
\end{equation}
Values of \(q\) greater than 2 result in decreasing \(j\) profiles, which render disks unstable to an axisymmetric instability discovered previously by Rayleigh \cite{Rayleigh17}.

The PPI has long been known to saturate into persistent
non-axisymmetric configurations of orbiting lumps\footnote{These are
  sometimes referred to as ``planets'', although they are not
  self-gravitating.} \cite{Hawley87}. Numerical relativity simulations
have shown that the PPI can occur in self-gravitating disks around
non-spinning BHs \cite{Korobkin11,Kiuchi11}, and generate
gravitational radiation \cite{Kiuchi11}.

While the conditions under which PPI unstable disks can form
dynamically are unclear, in \cite{Bonnerot2015} the simulation of the
tidal disruption of a white dwarf by a supermassive BH, found that a
nearly axisymmetric torus formed that was unstable to the
PPI. This remnant disk had a small mass, and subsequent studies
concluded that the resulting GW signal would be weak
\cite{Nealon17,Toscani19}. The numerical relativity simulations
in~\cite{Lovelace2013} find that the disks with disk to black hole
mass of $\sim 20\%$ forming dynamically following a black
hole--neutron star star merger appear to be stable for the times
simulated. However, these are limited studies and the parameter space
of black hole-neutron stars is large, so that more studies of such
mergers are necessary to understand if there exist conditions under
which PPI unstable disks form in these systems. On the other hand, the
simulations of supermassive stellar collapse in~\cite{Uchida:2017qwn}
find that the BH disks arising in the process have properties that are
favorable for developing the PPI. The above demonstrate that there
exist potential channels for the dynamical formation of PPI-unstable
disks, and hence it is worth exploring their potential as
multimessenger sources with gravitational waves. To better understand how
time changing quadrupole moments in disks around black holes result in
GW emission, it is more efficient to start with BH-disk initial data and evolve these as a means to study many progenitors at once, instead of performing simulations of the dynamical
formation of such disks from different progenitors. This is particularly feasible
when the matter is modeled with a $\Gamma$-law equation of state, in
which case there is an inherent scale freedom to the set of equations
governing the evolution. This is the approach we will adopt in this
work.

It is worth noting that magnetized disks unstable to the
magneto-rotational instability (MRI) could suppress the PPI
growth~\cite{Bugli2017}. However, both MRI and PPI are exponential
instabilities, and both occur on the orbital timescale of the
disk. Therefore, if conditions are such that the PPI develops first,
then it is plausible that the PPI can grow and survive for longer
times. This is supported by the findings of \cite{Bugli2017} who
demonstrated that when the \(m = 1\) mode is initially excited, which
is possible in dynamical disk formation scenarios, the PPI dominates
the dynamics for several disk orbits before finally succumbing to the
MRI. Therefore, it is important to further study the PPI, and most
importantly the detectability of the multimessenger signatures of
BH-disk systems.

In the context of simulations in full general relativity,
self-gravitating disks around black holes were studied
in~\cite{Korobkin:2012gj} where the runaway instability was
investigated. With regards to the focus of this work, the PPI has been
explored for self-gravitating disks with constraint-satisfying and
equilibrium initial data only around {\it non-spinning} black holes
\cite{Korobkin11,Kiuchi11}. Simulations around spinning black holes,
but adopting constraint-violating, and non-equilibrium initial data
were performed in~\cite{Mewes:2015gma}. Here, we initiate a study of
the dynamics of the PPI in massive, equilibrium, self-gravitating
disks around {\it spinning} black holes adopting constraint-satisfying
and equilibrium initial data~\cite{Tsokaros2019}. Our focus is to determine whether the
spin of the BH alters the onset of the PPI or its saturation state in
any way, and what effect BH spin may have on the detectability
of the GW signal, as well as other observables. Our initial disks all
obey the same rotation law, have an approximately flat specific angular
momentum profile (\(q\simeq 2\)), and approximately the same mass. In this work we
focus on scenarios where the BH spins have dimensionless value of $0$
or $0.7$,\footnote{Which coincides closely with the spins of the BHs formed by collapsing stars at the mass-shedding limit as shown by \cite{Shibata2002,Shapiro2002}.} either aligned or anti-aligned with the disk orbital angular
momentum. In forthcoming work, we will explore the dynamics of
misaligned BH spins, where spin-orbit precession effects are expected.

We find that the dynamics of the PPI in our aligned and anti-aligned
spin simulations is similar to the non-spinning case, which has been
studied previously. However, keeping other quantities nearly fixed,
the presence of BH spin causes unavoidable differences in the
structure of the disks, which in our case appear as shifts in the
orbital frequencies of the disks, thereby affecting the dynamics. The
differing innermost stable circular orbit (ISCO) radii also affect the
accretion rates, so that the case with the smallest ISCO accretes an
order of magnitude more slowly than the others. This result holds in
the absence of magnetic fields and for the particular initial inner
disk edges we start with; we will explore these effects in future
work. Following saturation of the PPI, the GW emission is nearly
monochromatic, and is strong enough to be detectable at great
distances by Cosmic Explorer, DECIGO, and LISA. In particular, for
systems where the disks are \(\sim 10\%\) the mass of the central BH,
Cosmic Explorer could detect a \(10 M_\odot\) system at 200-320 Mpc,
DECIGO could detect a \(10^3 M_\odot\) system out to \(z\sim5\), and
LISA could detect BH-disk systems with mass ${\rm few} \times
10^5M_\odot$ out to cosmological redshift of $z\sim 1$. Assuming that
1\% of the accretion power escapes as radiation, e.g., powering jets,
we find electromagnetic bolometric luminosities of
\(\mathcal{O}(10^{51-52})\)[erg/s] could arise from these systems,
making PPI-unstable disks promising multimessenger candidates with GWs
for the next generation of GW observatories.

The remainder of this paper is organized as follows: In Section
\ref{section:methods} we describe our initial data, and our methods
both for generating the initial data and for the dynamical evolutions.
In Section \ref{section:results} we discuss the results of the
evolutions, comparing the growth and saturation behavior of the
instability for all three BH spin states, as well as the GW signals,
and the properties of potential electromagnetic radiation. We then
scale our results to a number of astrophysically relevant regimes, and
analyze the detectability of such systems by LIGO, Cosmic Explorer,
DECIGO, and LISA. We conclude in Section \ref{section:discuss} with a
summary of our findings and discussion of future work.

Unless otherwise stated, all quantities are expressed in geometrized
units where \(G = c = 1\). Throughout \(M\) designates the
Christodoulou mass \cite{Christodoulou1970} of the central BH.

\section{Methods\label{section:methods}}

Our methods for generating initial data and for performing evolutions
have been described in detail elsewhere. Here we briefly summarize
these methods, pointing the reader to the appropriate references, and
list the properties of our initial data.

\subsection{Initial Data\label{section:id}}

Self-gravitating disks in equilibrium are computed using the COCAL
code and the techniques described in \cite{Tsokaros2019}. In
particular the complete initial value problem is solved for the full
spacetime metric including the conformal geometry. The Einstein
equations are written in an elliptic form and their solution is
obtained through the Komatsu-Eriguchi-Hachisu scheme \cite{KEH1989}
for black holes \cite{Tsokaros2007}.  

\begin{figure}[!htb]
\includegraphics{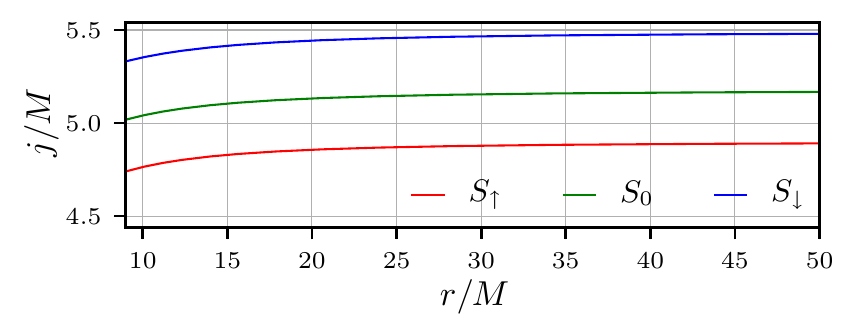}
\caption{Specific orbital angular momentum (\(j\)) on the equatorial plane as a 
function of cylindrical radius on the equatorial plane.}\label{fig:specific_j}
\end{figure}

\renewcommand{\tabcolsep}{1pt}
\begin{table}[!hbt]
\caption{Properties of the initial data for the three simulations:
  \(a/M\) is the dimensionless BH spin parameter whose sign implies
  whether the BH spin is aligned (+) or anti-aligned (-) with the disk
  orbital angular momentum; \(M_\mathrm{ADM}\) is the ADM mass of the
  spacetime; \(r_\mathrm{ISCO}\) is the cylindrical radius of the
  vacuum Kerr ISCO with the same BH Christodoulou mass and dimensionless spin;
  \(r_\mathrm{Inner},\ r_c,\ r_\mathrm{Outer}\) are the equatorial
  cylindrical radii of the disk inner edge, maximum density, and outer
  edge, respectively; \(\Omega_c\) is the orbital frequency at
  \(r_c\); and \(M_\mathrm{disk}\) is the total disk rest-mass.
}\label{tab:initial_data} \def\arraystretch{1.5}
\begin{tabular}{ccccccccc}
  \hline\hline
Label \ & \bfseries \(\frac{a}{M}\) & \bfseries \(\frac{M_\mathrm{ADM}}{M}\) & \bfseries \(\frac{r_\mathrm{ISCO}}{M}\) & \bfseries \(\frac{r_\mathrm{Inner}}{M}\) & \bfseries \(\frac{r_c}{M}\) & \bfseries \(\frac{r_\mathrm{Outer}}{M}\) & \bfseries {\(M\Omega_c\)} & \bfseries {\(\frac{M_\mathrm{disk}}{M}\)} \\ 
\hline \Su{}\ & 0.7 & 1.13 & 3.39 & 9.00 & 15.6 & 31.7 & \(1.61 \times 10^{-2}\) & 0.12 \\
 \So{}\ & 0.0 & 1.14 & 6.00 & 9.00 & 16.9 & 35.0 & \(1.47 \times 10^{-2}\) & 0.135 \\
 \Sd{}\ & -0.7 & 1.14 & 8.14 & 9.00 & 18.9 & 38.9 & \(1.23 \times 10^{-2}\) & 0.13\\
 \hline\hline
\end{tabular}
\end{table}

Our initial data of self-gravitating disks onto black holes correspond
to three BH spin-states, which we will refer to as \So{} for the
non-spinning case, and \Su{} (\Sd{}) for the case with black hole spin
aligned (anti-aligned) with the disk orbital angular momentum. Table
\ref{tab:initial_data} summarizes the parameters for all cases. Aside
from the BH spin, the BH Christodoulou mass (\(M\)) and the disk inner
edge were kept constant, while the rest of the quantities were
determined by keeping the disk rest-mass the same to within $\sim
10\%$ from the non-spinning case. Also, the ratio of the disk
rest-mass to the BH mass was chosen to be $\sim 0.1$.  The initial
spacetime closely corresponds to the Kerr metric, with distortions due
to the presence of the self-gravitating torus.  The disks are modeled
as perfect fluids obeying a polytropic equation of state
\begin{equation}
P = k\rho_0^{\Gamma},
\end{equation}
where \(\Gamma = 4/3\), appropriate for a radiation-pressure dominated
gas. The disk inner edge radius was chosen to be at least 10\% greater
than the corresponding vacuum Kerr ISCO radius in all cases.

The crucial ingredient for the PPI is the differential rotating law of
the disk. As in \cite{Tsokaros2019} we assume that the relativistic
specific angular momentum $j=u^tu_\phi$ profile is given by
$j(\Omega)= A^2(B_0-\Omega)$ with $A=0.1$ and $B_0$ a constant that is
evaluated during the iteration scheme. This choice leads to a nearly
constant ($j\sim r^{0.01}$) angular momentum profile as can be seen in
Fig. \ref{fig:specific_j} that renders the disk unstable to the PPI.  In
terms of the Newtonian Eq. (\ref{eq:newj}) we have $q\approx 2$.

Solutions were then generated satisfying these conditions for each of
the three BH spin states and solving the complete initial value
problem~\cite{Tsokaros2019}. The resulting disks differ in their radii
of maximum density, which are smaller for more positive spin
values. This is responsible for the increase of \(M \Omega_c\) seen in
Table \ref{tab:initial_data}, as the values of \(\Omega_c\) computed
for circular geodesics at the corresponding radii in the Kerr metric
agree closely with the orbital frequencies of our initial data. The
differences in orbital period are small but not insignificant, and 
end up affecting the dynamics of the disks, as we discuss in Section
\ref{section:results}.

\subsection{Evolution\label{section:evolution}}
The BH-disk systems were evolved with the Illinois dynamical spacetime,
general relativistic magnetohydrodynamics adaptive-mesh-refinement
code~\cite{Duez2005,Etienne2010,Etienne2012}. Built within the
Cactus/Carpet infrastructure \cite{Cactus,Carpet}, this code is the
basis of the publicly available counterpart in the Einstein toolkit
\cite{Etienne2015}. The spacetime metric is evolved by solving the
Baumgarte-Shapiro-Shibata-Nakamura (BSSN) equations
\cite{Shibata1995,Baumgarte1998} using the moving-puncture gauge
conditions \cite{Baker2006,Campanelli2006}, with the shift vector
equation cast into first-order form (see e.g. \cite{Hinder2013}).  The
fluid is evolved using a $\Gamma$-law equation of state,
$P=(\Gamma-1)\rho_0\epsilon$, where $\Gamma=4/3$, $\rho_0$ is the
rest-mass density, and $\epsilon$ the internal specific energy.

\subsubsection{Grid hierarchy}
The evolution grid hierarchy consists of nested cubes, demarcating 11
concentric refinement levels. Here we will refer to the levels by
their index, $n$, where $n=1$ corresponds to the finest level and
$n=11$ the coarsest. The finest level half-side length is set to $r_1
= 2.19M$, and the first three are then $r_n = 2^{(n-1)} r_1$ $(n \le
3)$. The remaining levels have half-side lengths $r_n = 2^{n} r_1$ $(n
> 3)$. The physical extent of levels $n\ge4$ is increased by the extra
factor of 2 to provide high resolution over the extended area of
the disk. Thus, the outermost level has a half-side length of $2250M$.

We set the spatial resolution on the finest level to $dx_1 =
M/25.6$. Each subsequent refinement level has half the resolution of
the previous. Therefore, the resolution of refinement level $n$ is
given by $dx_n = 2^{(n-1)}dx_1$. We adopt Cartesian coordinates, and
equal spatial resolution is chosen for the $x$, $y$, and $z$
directions, without imposing any symmetries on the grid.

\subsubsection{Diagnostics}

During the evolution, we monitor the normalized Hamiltonian and
momentum constraints calculated by Eqs.~(40)--(43) of~\cite{eflstb08}.

The growth of the unstable density modes was tracked by evaluating the
following integral at regular time intervals
\begin{equation}\label{eq:mode_def}
C_m = \int{\sqrt{-g}\mathrm{d}^3x u^0\rho_0 e^{im\phi}},
\end{equation}
which provides a measure of the non-axisymmetric rest-mass density
modes that develop (see
e.g.~\cite{Paschalidis:2015mla,East:2015vix,East:2016zvv}). Here $g$
is the determinant of the spacetime metric, $u^0$ the 0 component of
the fluid four-velocity, and $\phi$ the azimuthal angle.

GWs are extracted using the Newman-Penrose Weyl scalar
$\psi_4$ at various extraction radii. We decompose $\psi_4$ into
$s=-2$ spin-weighted spherical harmonics up to and including $l=3$
modes. The GW polarizations $h_+$ and $h_\times$ for each mode are computed
by integrating the corresponding mode of $\psi_4$ twice with time
using the fixed frequency integration technique described
in~\cite{Reisswig:2010di}.

\section{Results\label{section:results}}

We performed two types of evolutions of our initial data.  First, we
simulated the systems in the Cowling approximation (where the
spacetime metric is held fixed) to study the early growth of the PPI
before it turns non-linear and to corroborate that the characteristics
of the instability match those of the PPI. Then we turned on the
spacetime evolution and evolved through the instability non-linear
growth, saturation and steady state. The results of these simulations
are described in the following sections.

\subsection{Cowling Approximation\label{section:cowling_growth}}
Analytical studies of the PPI looked at its early growth phase for
tori in stationary spacetimes, such as in \cite{Kojima86b} where a
Schwarzschild background was assumed. To make contact with these
earlier works, we evolved our initial data using the Cowling
approximation.  While not identical to the disks in \cite{Kojima86b}
(the background spacetime is not precisely Kerr due to the disks'
self-gravity), this allows us to qualitatively compare the early growth
in our simulations to analytical expectations, without any of the
effects of back-reaction onto the spacetime. More importantly, fixing
the background spacetime metric makes the spacetime coordinates
well-defined as ``Kerr-Schild''-like coordinates. Through the Cowling
approximation evolutions the early-time PPI growth rates can be
estimated.

\begin{figure}[!htb]
\includegraphics{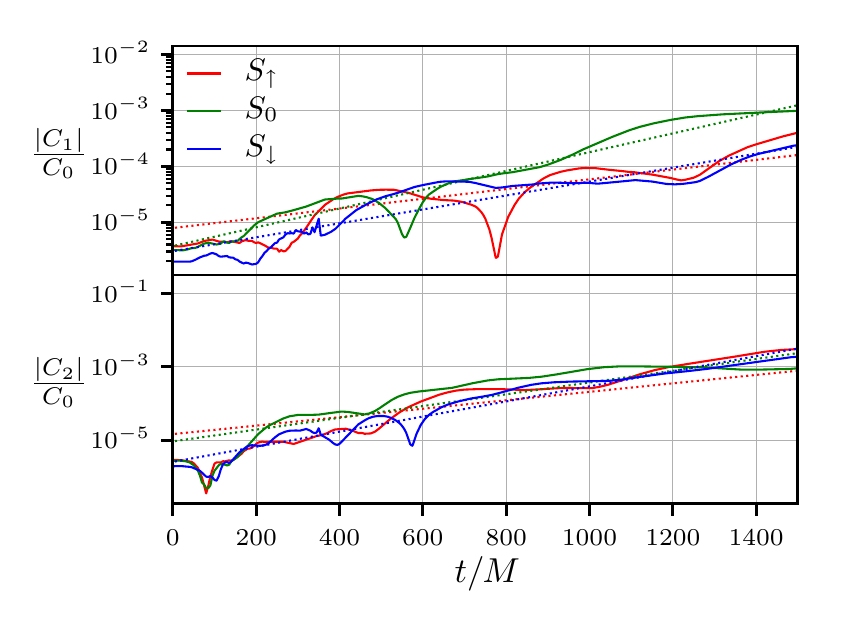}
\caption{Early amplitude growth of the non-axisymmetric density modes for \(m=1\) and \(m=2\), evolved with the spacetime background held fixed. A linear fit to each curve is plotted (dotted lines), which indicates the exponential growth rate of the modes.}\label{fig:dens_modes_log_static}
\end{figure}

\begin{table}[!hbt]
\caption{Exponential growth rates of \(m=1\) modes in the fixed spacetime evolutions. \(\sigma\) is the complex mode frequency, defined in Equation \ref{eq:mode_form}. The factor \(r_\mathrm{Inner}/r_c\) was used in \cite{Kojima86b} to parameterize disk geometries, and is included here to show how our disks relate to those studied in that paper.}\label{tab:growth_rates}
\def\arraystretch{1.5}
\begin{tabular}{ccc}
  \hline\hline
  Label & \bfseries {\(r_\mathrm{Inner}/r_c\)} & \bfseries {\(\operatorname{Im}(\sigma)/\Omega_c\)} \\ 
  \hline
 \Su{} \ & \(5.76 \times 10^{-1}\) & \(1.25 \times 10^{-1}\) \\
 \So{} \ & \(5.34 \times 10^{-1}\) & \(2.64 \times 10^{-1}\) \\
 \Sd{} \ & \(4.76 \times 10^{-1}\) & \(2.34 \times 10^{-1}\) \\
 \hline\hline
\end{tabular}
\end{table}

\begin{figure*}[!htb]
\includegraphics{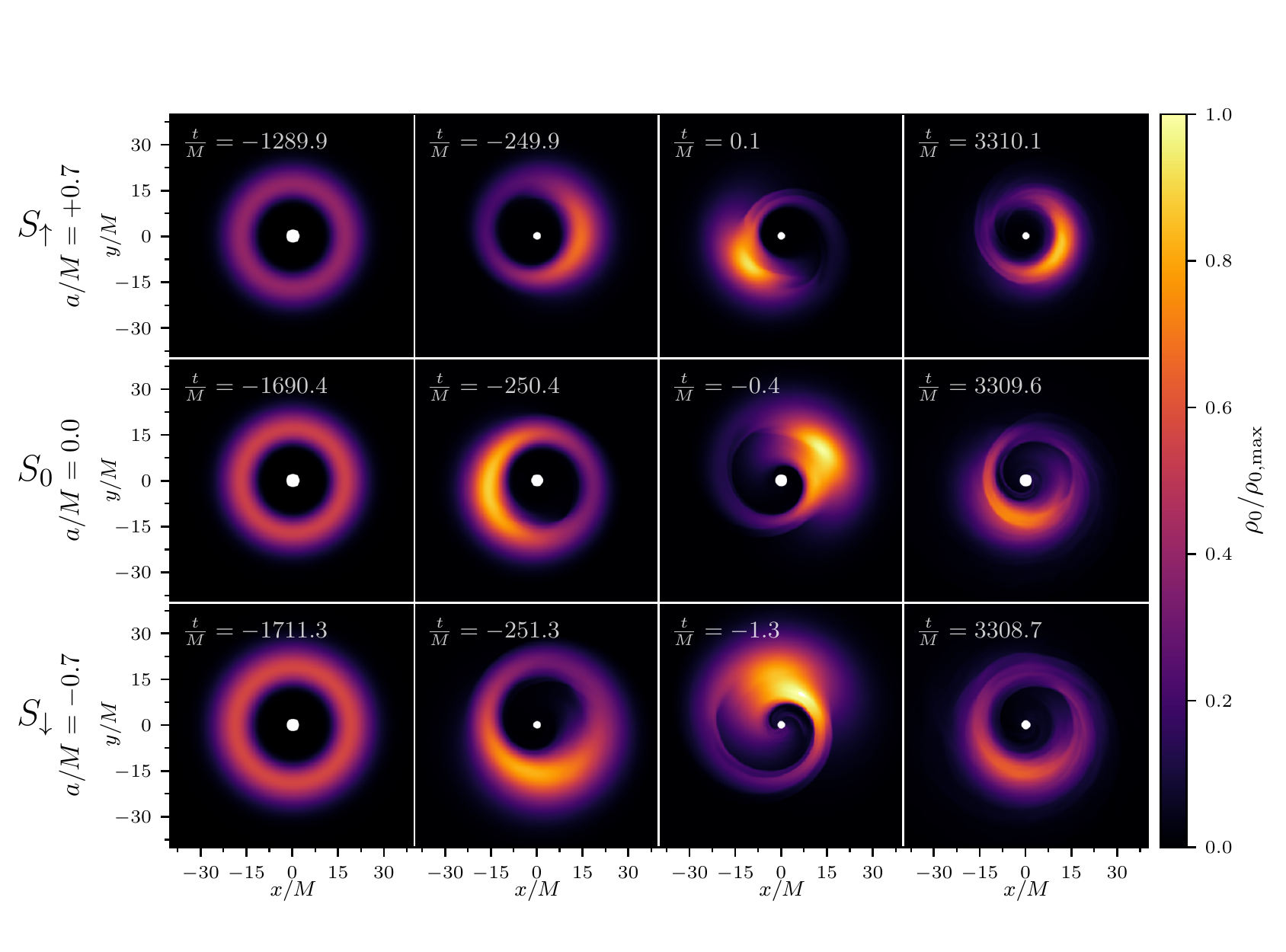}
\caption{Snapshots of xy-plane cross-sections of the rest-mass density
  normalized to its maximum value over the entire evolution. Each row
  corresponds to a single simulation (labeled on the left \Su{},
  \So{}, \Sd{}), with snapshots arranged chronologically from left to
  right. The initial time is negative because we have shifted the time
  such that \(t=0\) is the time of the PPI saturation (defined to be
  the time at which the amplitude of the \(m=1\) mode becomes maximum
  for the first time), which approximately coincides with the third
  column snapshots. Apparent horizon regions are filled in white, and
  the snapshots are centered on the BH coordinate centroids. (The
  change in size of the horizons between the first and second
  snapshots of each simulation is due to coordinate
  relaxation.)}\label{fig:rho_b_frames_xy}
\end{figure*}

Perturbations seeded due to finite-resolution excite all modes to a
small degree. At early times the modes grow exponentially until only
the fastest-growing ones dominate. For the tori geometries we
simulated, analytical studies predict the dominant modes to be \(m=1\)
and \(m=2\)~\cite{Kojima86a}. This was observed in our simulations as
well. In Figure~\ref{fig:dens_modes_log_static} we plot the \(m=1\)
and \(m=2\) mode amplitudes for all three disks as computed based on
Eq.~\eqref{eq:mode_def}. The growth of these modes approximately
follows an exponential trend, as shown by the dotted lines in the
figure. The estimated exponential growth rates are reported in
Table~\ref{tab:growth_rates}, and are of the order
\(\mathcal{O}(0.1)\) when normalized to $\Omega_c$. While a direct
comparison with the disks onto Schwarzschild black holes in
\cite{Kojima86b} is not possible, because we have different disks and
spacetimes, we find broad agreement with the rates calculated
semi-analytically in \cite{Kojima86b} when comparing models with
approximately the same \(r_\mathrm{Inner}/r_c\) - the way
\cite{Kojima86b} parametrized the disks. In addition, we find
qualitative agreement with \cite{Kojima86b} in that the growth rate is
exponential, and that the low-\(m\) modes dominate, supporting the
conclusion that the instability that developes in our simulations is
the PPI, as expected from the specific angular momentum profile of the
disk.

The self-gravitation of our tori makes the Cowling approximation
unsuitable for studying the dynamics through saturation, thus we also
evolved them in dynamical spacetime. We turn next to our dynamical
spacetime study, which showcases the full dynamics of these systems
from early growth until long after saturation.

\subsection{Dynamical Spacetime\label{section:growth_sat}}

When the disks are evolved in full general relativity, all three
undergo violent non-axisymmetric instabilities. As shown in the first
three columns of Figure~\ref{fig:rho_b_frames_xy}, all three disks
develop non-axisymmetric density modes that grow quickly to saturation
over a few orbits. Shocks develop during the development of the
instability, which redistribute angular momentum until finally the
density pattern saturates with the $m=1$ mode dominating the subsequent
evolution. The right-most column of the figure contains snapshots of
the disks long after saturation, which show that the density mode
pattern still persists.

\begin{figure}[!htb]
\includegraphics{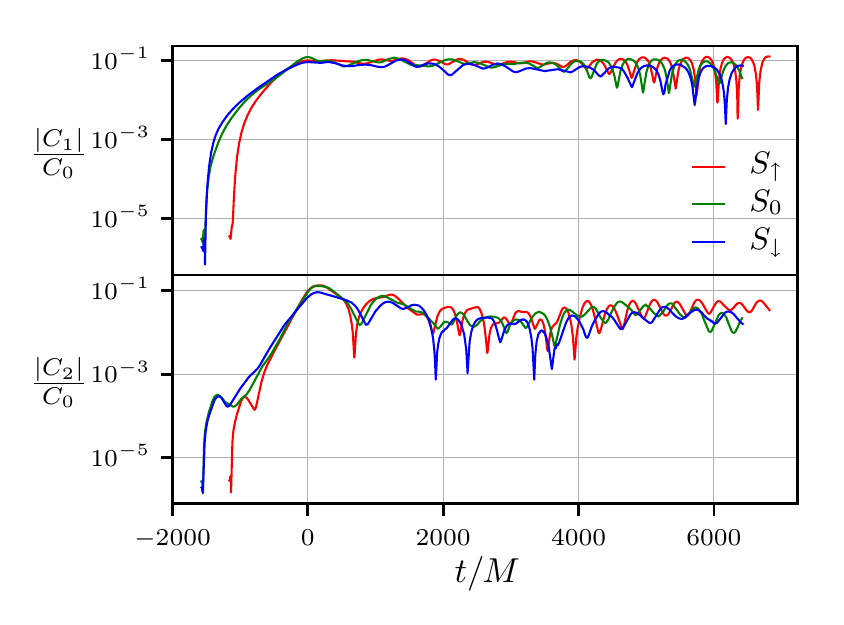}
\caption{Full dynamical evolution of the amplitude of the non-axisymmetric \(m=1\)
  and \(m=2\) density modes, normalized by the \(m=0\) mode
  amplitude.}\label{fig:dens_modes_log}
\end{figure}

Density mode amplitudes are shown in
Figure~\ref{fig:dens_modes_log}. As in the Cowling approximation
evolutions, we find that the fastest modes are still \(m=1\) and
\(m=2\), which are the ones plotted. The figure reveals that the
normalized non-axisymmetric \(m=1\) and \(m=2\) density modes for all
three disks saturate near the same values, but the $m=1$ mode
dominates. The PPI growth as measured by the $C_m$ is still
approximately exponential in coordinate time before saturation. In
\cite{Korobkin11} the PPI growth rate was reported to be slightly
greater in a dynamical spacetime than for a fixed spacetime.  Although
statements based on coordinate time are gauge-dependent in evolutions
where the spacetime and coordinates are dynamical, our results show
qualitative agreement with this previous finding: the instabilities
grow more quickly in the dynamical spacetime evolutions.

\begin{figure}[!htb]
\includegraphics{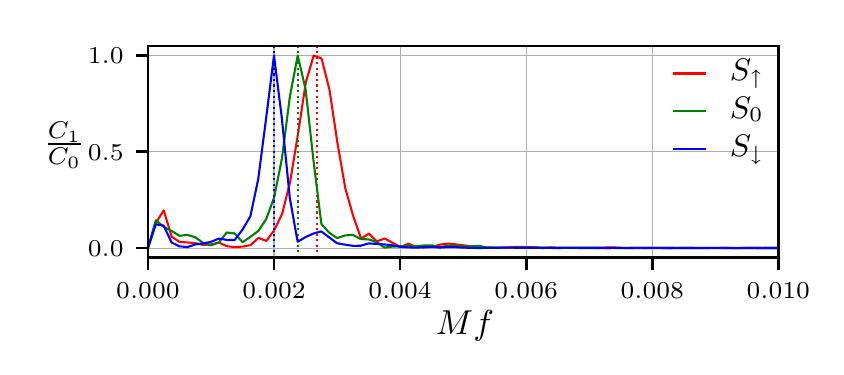}
\caption{Normalized spectra of the \(m=1\) non-axisymmetric density mode for the three disks. The dominant frequency of the \(m=2\) modes is twice that of \(m=1\) in each case.}\label{fig:dens_modes_freq}
\end{figure}

\begin{figure}[!htb]
\includegraphics{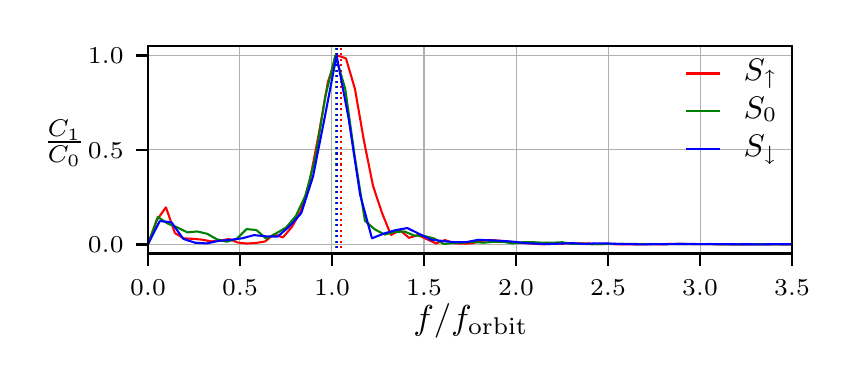}
\caption{Spectra of the m=1 mode from
  Figure~\ref{fig:dens_modes_freq}, plotted relative to each disk's
  respective initial orbital frequency at maximum density. The plot
  shows reveals that the density modes among different cases are in
  close agreement agreement.}\label{fig:dens_modes_freq_orb}
\end{figure}

In Figure~\ref{fig:dens_modes_freq} we compare the spectra of the
modes, which are similar across the different cases, with the only
significant difference being the location of the peak frequency. We
find that the peak frequencies correlate closely with the orbital
frequencies at maximum density of the initial data. When plotted
relative to each disk's respective orbital frequency, as in
Figure~\ref{fig:dens_modes_freq_orb}, the spectra are nearly
identical. The spectra for the \(m=2\) modes are less clean, but their
peak frequencies are double that of the \(m=1\) modes in each case, as
anticipated when the dynamics is driven by an \(m=1\)
mode~\cite{East:2015vix}, and therefore they are nearly equal for the different cases when
scaled by orbital frequency. We also find that higher-\(m\) modes are
excited but are orders of magnitude weaker than the \(m=1\) and
\(m=2\) modes.

The above results suggest that in the case of aligned or anti-aligned
BH spins, we do not observe any significant difference from the PPI's
established behavior around non-spinning black holes. The only change
is a frequency shift that matches the orbital frequency shift between
the initial data for each case. As discussed in Sec.~\ref{section:id},
these orbital frequency shifts originate from slight differences in
the initial data and are driven by the BH spin. If we were to alter
the disk parameters so that the initial orbital frequencies match, we
would have to choose disks with different values for
\(r_\mathrm{inner}\) or \(M_\mathrm{disk}\) or change the rotation
law, which could alter the PPI in other ways. Here we chose to fix
\(R_\mathrm{inner}\) and approximately \(M_\mathrm{disk}\) instead, to
focus on the effects of the BH spin.

Ultimately, the effect of BH spin is both indirect and
unavoidable. Although there seems to be no direct change in the nature
of the PPI due to spin (at least not for spins up to \(a/M=0.7\)), the different spin states still force disks to
assume different structures, which alter the PPI in a predictable
way. Therefore, spin is an important parameter to consider when
exploring the range of dynamics of possible PPI-unstable BH-disk
systems, and predicting their gravitational-wave signatures.

As mentioned above, while the disk density modes are useful for
understanding the character of the instability, they suffer from gauge
ambiguities. We can lift these ambiguities by studying the instability
through the gravitational radiation instead, which can be extracted
unambiguously.

\subsection{Gravitational Wave Signal\label{section:gw_signal}}

\begin{figure}[!htb]
\includegraphics{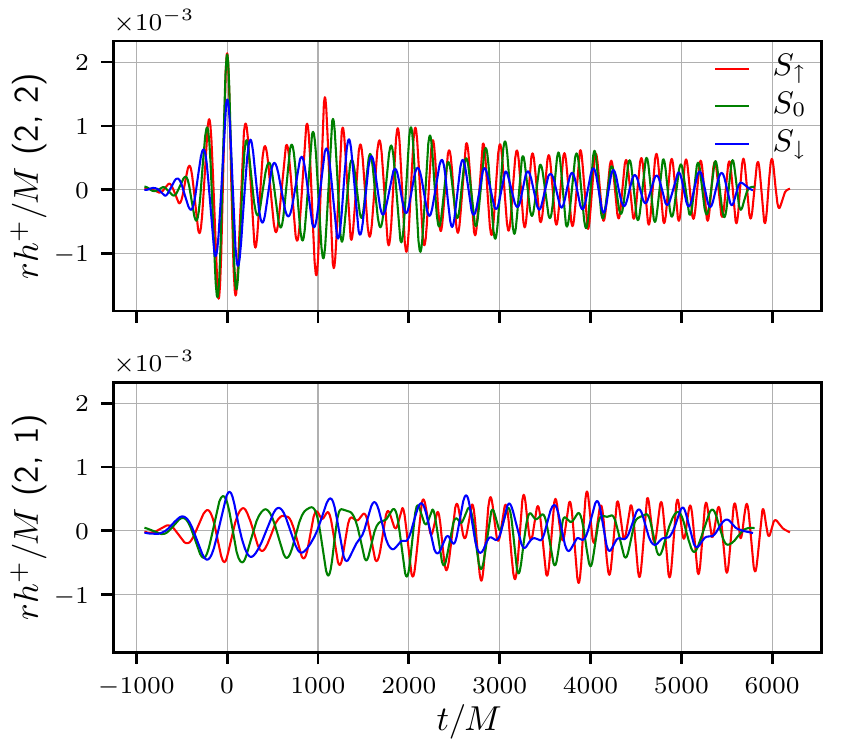}
\caption{\label{fig:comp_strain}Strain waveforms for the \(l=2\),
  \(m=2\) (top), and \(m=1\) (bottom) radiation multipoles. To ease
  comparison, phases have been rotated to align at maximum amplitude
  of the \(l=2\), \(m=2\) mode.}
\end{figure}

In Fig.~\ref{fig:comp_strain} we compare the \(l=2\), \(m=2\) and
\(l=2\), \(m=1\) multipole moments of the gravitational radiation. All
three cases exhibit an initial burst corresponding to the saturation
of the instability, and then a relaxation to a quasi-monochromatic
signal of lower amplitude. The $l=2$, $m=2$ plot in
Fig.~\ref{fig:comp_strain} exhibits a noticeable difference between
the peak amplitudes of the signal from disks \Su{} and \So{}, and disk
\Sd{}.  What is the reason for the difference in signal amplitude? The
quadrupole formula provides insight. To a rough approximation we can
model the BH-$m=1$ mode-dominated disk system as a pair of orbiting
point masses, one representing the black hole, and one representing
the displaced center-of-mass of the dominant non-axisymmetric mode. In
this model, we assume that the effective reduced mass \(\mu\) is
essentially the disk mass due to the small mass ratio. Then, assuming
that the orbital separation \(r_c\) and reduced mass \(\mu\) change
slowly relative to \(\Omega_c\), the quadrupole formula predicts the
strain signal\footnote{With the real and complex parts representing
  \(+\) and \(\times\) polarizations, respectively} for viewpoints in
the orbital plane to be (for derivation, see \cite{Lai1994}; see
also~\cite{Paschalidis:2009zz}):

\begin{equation}\label{eq:quad_signal}
rh = 4 r_c^2 \mu \Omega_c^2 e^{2 i \Omega_c t},
\end{equation}
where \(\Omega_c\) is the orbital frequency, which we also take from
Table~\ref{tab:initial_data}. All three disks have nearly identical
profiles of \(|C_m|/C_0\) for \(m=1\) and \(m=2\), with the
higher-\(m\) having much smaller amplitudes, and after \(\sim 2500 M\)
accretion has resulted in similar disk masses. Therefore, in the model
we can assume they each have the same values of \(\mu\) (the orbital 
motion of the effective point mass can account for the phase rotation of \(C_m\), since they are observed to have frequencies proportional to \(m\)).
Then we can use the values of \(\Omega_c\) and \(r_c\) from
Table~\ref{tab:initial_data} to compute the expected amplitude ratios
between the three disks. We find that the amplitude of \So{} should be
\(1.14\times\) that of \Sd{}, and the amplitude of \Su{} should be
\(1.21\times\) that of \Sd{}, which is in broad agreement with what we
observe. Hence, the difference in GW amplitude can be accounted
for as yet another effect of the shift in disk orbital frequency due
to the different BH spin states.

\begin{figure}[!htb]
\includegraphics{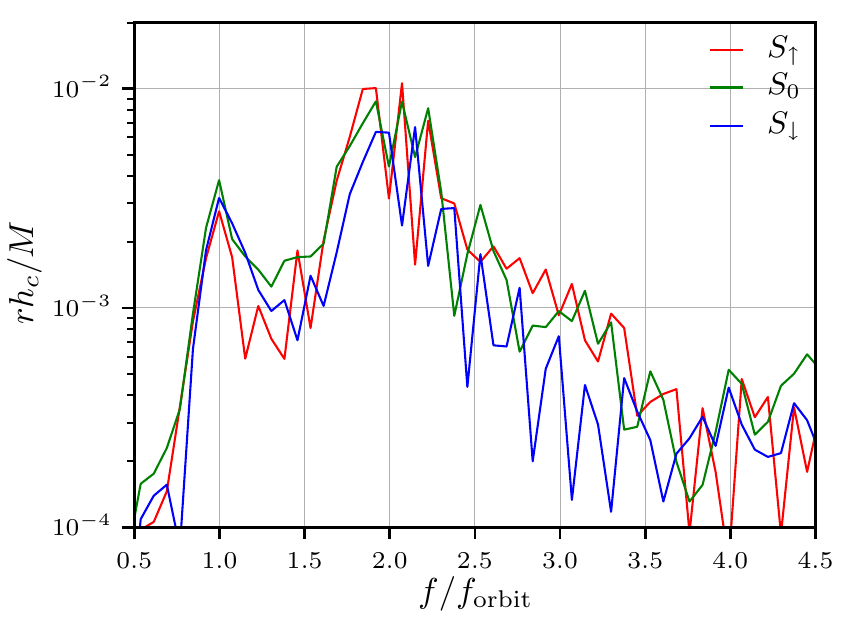}
\caption{\label{fig:comp_char_strain} Frequency-domain comparisons of the polarization-averaged characteristic strain (defined in (\ref{eq:char_str_def},\ref{eq:pol_avg_def})) of the GWs from the three disks, with respect to the orbital frequency. All radiation multipoles are combined with appropriate spin-weight spherical harmonics, assuming a viewing angle of \(\theta = \pi/2.34\) away from the orbital axis, which results in the observed amplitude of the \(l=2, m=2\) mode being equal to its direction-averaged value.}
\end{figure}

To quantify the frequency-domain behavior we calculate the
\textit{characteristic strain} (see \cite{Moore2014}), which is
defined only over positive frequencies as,

\begin{equation}\label{eq:char_str_def}
h_c = 2f|\tilde{h}^\mathrm{res}|,
\end{equation}
where \(\tilde{h}^\mathrm{res}\) is conventionally taken to be the Fourier transform of the interferometer response to the incoming strain waveform \(h\). However, in this work we will consider multiple detectors with different response functions, so we instead choose,

\begin{equation}\label{eq:pol_avg_def}
\tilde{h}^\mathrm{res} = \sqrt{\frac{|\tilde{h}_+|^2 + |\tilde{h}_\times|^2}{2}},
\end{equation}
where \(\tilde{h}_+, \tilde{h}_\times\) are the Fourier transform of
the \(+\) and \(\times\) polarizations of the incoming signal. For the
remainder of this paper we will take \(h_c\) to be the
``polarization-averaged'' characteristic strain, defined through
Eqs.~\eqref{eq:char_str_def} and~\eqref{eq:pol_avg_def}. 

Figure~\ref{fig:comp_char_strain} compares the characteristic strain
of each waveform for an angle between the line of sight and the
orbital plane that corresponds to a value such that $_{-2}Y^2_2$
equals its angle-averaged value. All extracted radiation multipoles up
to and including $l=3$ are used in the calculation. Once again,
measured relative to the orbital frequencies the spectral peaks of the
3 different cases align well with each other, with the dominant peaks
occurring at \(f_\mathrm{orbit}\) and \(2f_\mathrm{orbit}\) for the
\(m=1\) and \(m=2\) modes, respectively. This figure demonstrates in a
gauge-independent way the results we found using the density modes in
the previous section: the dominant non-axisymmetric modes are the
$m=1$ and $m=2$.

\subsection{Accretion and possible electromagnetic counterparts\label{section:accretion}}

In addition to GWs, BH-disk systems are likely to emit electromagnetic
(EM) radiation because of accretion. As the disk undergoes dynamical
relaxation, shocks within the disk redistribute angular momentum,
which allows accretion to proceed. Associated with this accretion
mechanism bright, electromagnetic counterparts are possible. While
there is no source of (effective) viscosity in our simulations, if net
poloidal magnetic flux is accreted onto the black hole at the rate
found in our simulations it would power
jets~\cite{Paschalidis:2014qra}.  In cases where a viscous dissipation
mechanism is involved emission is expected to arise locally as
gravitational binding energy is released when matter is gradually
transported to circular orbits closer to the BH. If the disks become
dense and hot enough they can also generate copious neutrino emission,
but this is more relevant to stellar mass systems. The power available
for EM emission is usually taken to be proportional to the accretion
power. Under this assumption we can therefore expect the luminosity of
the disk to obey
\begin{equation}\label{eq:EMluminosity}
L_\mathrm{EM} = \epsilon\dot{M}_\mathrm{disk}c^2,
\end{equation}
where $\dot{M}_\mathrm{disk}$ is the rest-mass accretion rate, and
$\epsilon$ is the efficiency for converting accretion power to EM
luminosity. For geometrically thin disks in the Kerr metric the
difference in binding energy between infinity and ISCO allows maximum
possible values of \(\epsilon\) up to 40\%, depending on BH spin, and
in astrophysically realistic settings \(\epsilon\) is typically
estimated to be \(\approx\)10\% \cite{Shapiro1983}. For thick disks
around BHs arising from magnetized black hole-neutron star mergers,
binary neutron star mergers or supermassive stellar collapse Poynting
dominated jet power satisfies $\epsilon\simeq
0.1$\%-$0.5$\%~\cite{Paschalidis:2014qra,Ruiz:2016rai,Sun2017,Ruiz:2019ezy}.
Here we will adopt a nominal value of $\epsilon=1$\%, but it is
important to keep in mind that it is possible that $\epsilon \ll
1\%$. We will test this assumption in a forthcoming work where we
treat the effects of magnetic fields. Note that even if magnetic
fields allow the PPI to operate for a few orbits~\cite{Bugli2017},
  the GW signal would likely be accompanied by a magnetically powered
  jet.

\begin{figure}[!htb]
\includegraphics{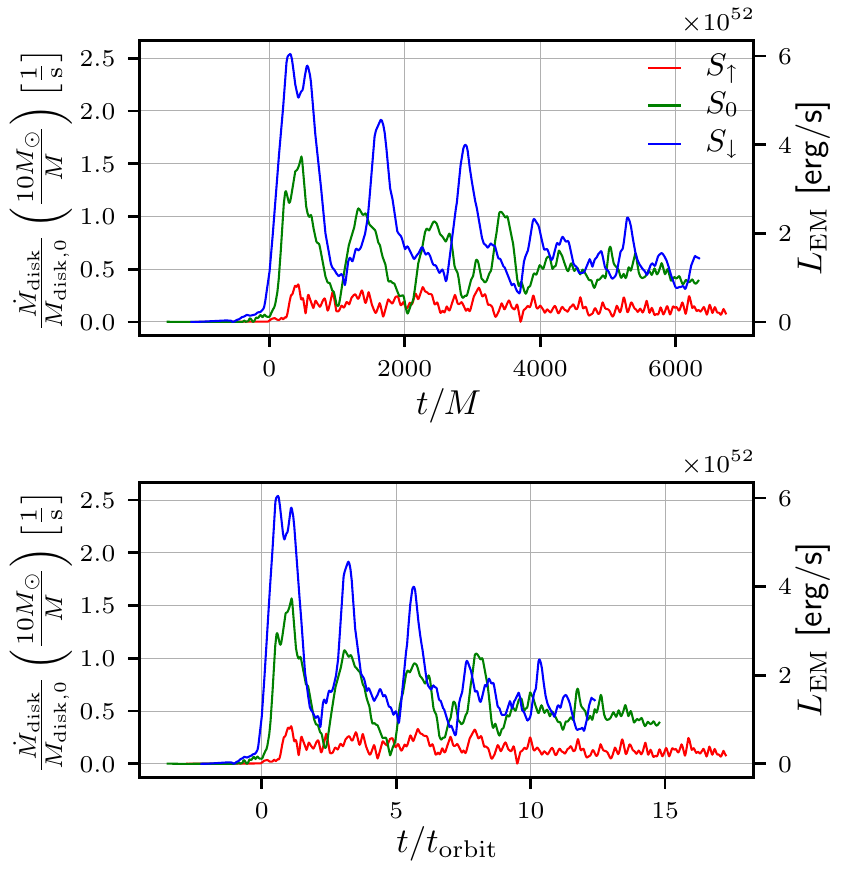}
\caption{\label{fig:accretion_comp}
  Rest-mass accretion timescales and estimated EM luminosities for the
  cases in our study. The accretion timescales (inverse accretion
  rates) are marked on the left y-axis, and scale with \(1/M\) as indicated by the axis label.
  The right y-axis shows the estimated bolometric luminosities
  (which do not scale with mass) using Eq.~\eqref{eq:EMluminosity}
  with \(\epsilon=1\%\). The bolometric luminosities of \So{} and
  \Sd{} are \(\mathcal{O}(10^{52})\)[erg/s], while \Su{} only achieves
  \(\mathcal{O}(10^{51})\)[erg/s]. In the top panel the coordinate time is
  normalized to the BH mass. In the bottom panel the time is scaled by the
  initial orbital period at maximum density, showing that the period
  of the accretion spikes scales with the orbital period for the three
  simulations.}
\end{figure}

In Figure~\ref{fig:accretion_comp} the rest-mass accretion rate is
plotted for all three cases. Disks \So{} and \Sd{} display repeated
spikes in their accretion rates, the periods of which scale
approximately with \(t_\mathrm{orb}\), as shown in panel (b). As time
goes on the accretion rates of \So{} and \Sd{} become less volatile,
so the spiking is likely a transient effect associated with the
relaxation of the initial data into the PPI saturation phase. The root
cause of this spiking is unclear, and it is unknown whether these
features are generic to PPI-unstable disks or unique to the specific
initial configurations we evolved.

\begin{figure*}[!htb]
\includegraphics{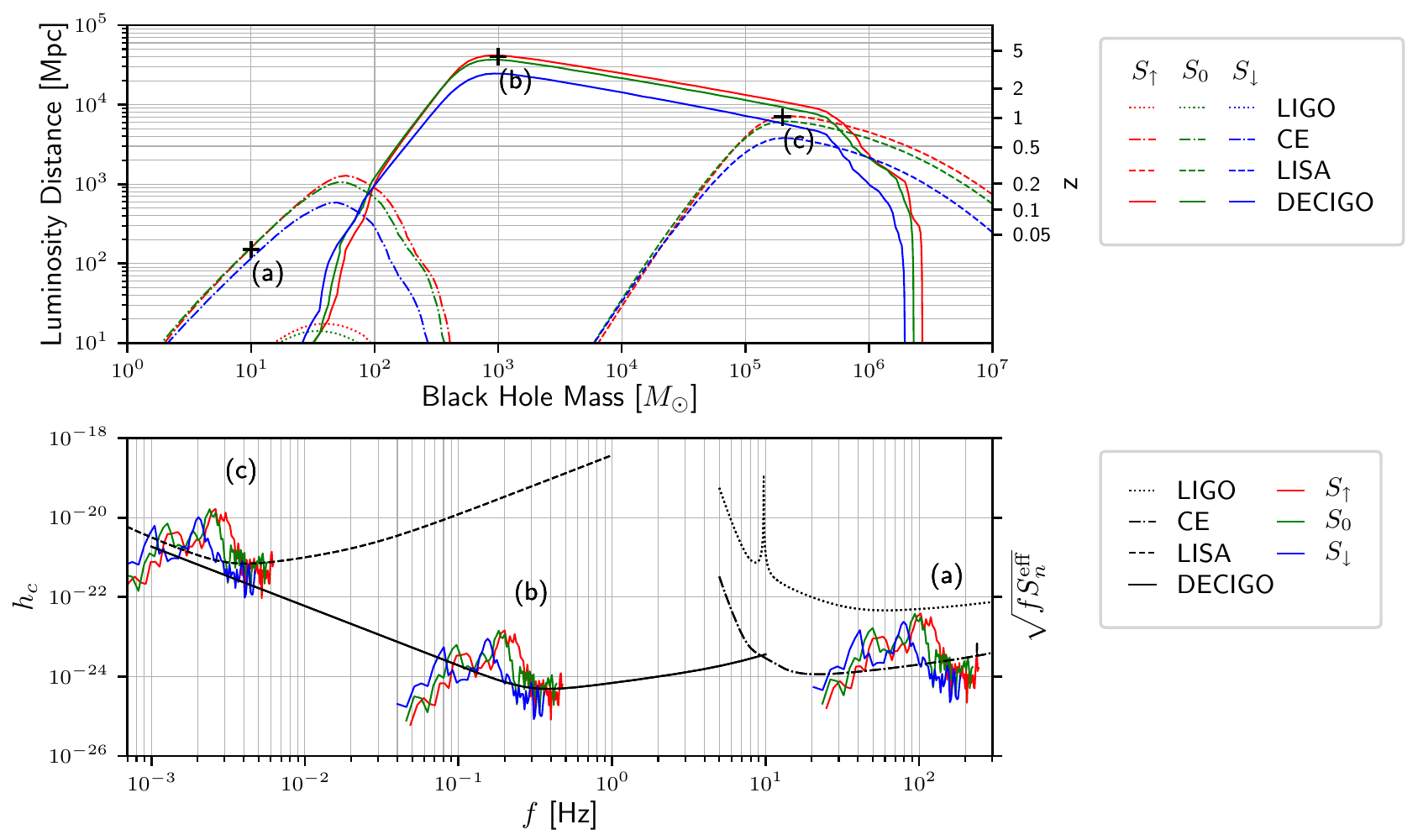}
	\caption{\label{fig:detection_horizon} Top:
          Maximum detection distances for systems over a range of BH
          masses, assuming a detection threshold of SNR=8 (sky \&
          polarization averaged). The colors and line patterns
          correspond to the spin state and detector, as shown in the
          legend. Three hypothetical sources are marked by crosses,
          representing the most distant systems at three different BH
          masses that can be detected assuming the \Su{} waveform: (a)
          \(10 M_\odot\), \(150\) Mpc; (b) \(1000 M_\odot\), 40000 Mpc
          (\(z=4.3\)); (c) \(2\times10^5 M_\odot\), \(7000\) Mpc
          (\(z=1.02\)).  Bottom: Characteristic
          strain curves (see equations~\eqref{eq:char_str_def} and
          \eqref{eq:pol_avg_def})) for each spin state (colored lines)
          with overlaid the sky-averaged detector characteristic sensitivity
          curves for LIGO (dotted), Cosmic Explorer (dot-dashed),
          DECIGO (solid), and LISA (dashed). The labels by each set of
          characteristic strain curves indicate the BH mass and
          luminosity distance of the source, which are marked on the
          bottom panel. The area between the sensitivity and signal
          curves determines the SNR. In both plots, we consider only 
          the part of the signal after \(t = 1000 M\), as the earlier
          portion may be strongly influenced by transients arising from
          hydrodynamic relaxation of the initial data.}
\end{figure*}

After the accretion rates settle down, the disk in case \Su{} ends up
with a significantly suppressed rate relative to \So{} and \Sd{}. A
likely reason for this is that the ISCO is further away from the inner
edge of the disk in \Su{} than it is for \So{} and \Sd{} (for the
latter the ISCO nearly coincides with the inner edge). This difference
in accretion rate also translates to significant difference in
estimated bolometric luminosity. As shown by the right axis of
Fig.~\ref{fig:accretion_comp}, after the transient accretion spikes
die away, the luminosities of \So{} and \Sd{} are
\(\mathcal{O}(10^{52})\)[erg/s], while \Su{} is an order of magnitude
dimmer at \(\mathcal{O}(10^{51})\)[erg/s]. Even the dimmest of these
bolometric luminosities is high enough to be detectable over a large
distance, and makes such disks potentially promising sources of
electromagnetic radiation, provided the conversion efficiency to
observable frequencies is not too low. Note that for stellar mass
black holes it is possible that much of that power is in the form of
neutrinos instead, but we would still expect jets to arise if net
poloidal magnetic flux is accreted onto the black ho.e

The accretion rate is also important as the determiner of the disk
lifetime, and hence the time over which the system emits gravitational
radiation. Notably, the \Su{} accretion timescale is significantly
greater than that reported by \cite{Kiuchi11} for the strictly
non-spinning case. Scaling the rates reported by \cite{Kiuchi11} to a
10 \(M_\odot\) BH surrounded by a disk of \(\sim\)\%10 its mass, we
obtain an accretion timescale between $\sim 0.5$s and $\sim 4$s. This is
consistent with the \So{} timescale we find, which is about
2.5s. However, for \Su{}, the accretion timescale is about 10s (as can
be seen in Figure \ref{fig:accretion_comp}), which is significantly
longer. This improves the detection prospects of such BH-disk systems,
which are analyzed in the next section. However, we note that magnetic
fields should be accounted for to test if these accretion rates are
robust against the MRI. This will be the topic of future work of ours.

\subsection{Detectability of gravitational waves\label{section:detect}}

In the previous sections, all results were reported in terms of
dimensionless quantities natural for the system being considered. To
assess detectability it is necessary to give the systems a definite
physical scale. Both GR and the equations governing the \(\Gamma\)-law
perfect fluid scale with the system's total gravitational mass. Since
in our simulations we consider disks roughly \(10\%\) the mass of the
central BH, it is the BH mass that primarily determines the mass of
the system.  In this section we exploit the scale-invariance of our
simulations to apply our results to a broad range of masses and
astrophysical systems.

There are three mass ranges of astrophysical relevance.  On the more
massive side are BH-disk systems of \(10^3\)-\(10^6 M_\odot\).  It has
been shown that systems of such masses, with disks \(\approx10\)\%
\(M\), can be formed by collapsing super-massive stars (SMS)
\cite{Shibata2002,Shapiro2002,Shapiro2004,Sun2017,Uchida2017}. Such SMS collapses have been
conjectured to occur in the early Universe, providing seed BHs which
may grow into supermassive BHs
\cite{Loeb1994,Shapiro2003,Koushiappas2004,Shapiro2004a,Shapiro2005,Begelman2006,Lodato2006,Begelman2009},
the early appearance of which at \(z\sim7\) is challenging to explain
(see reviews \cite{Haiman2012,Latif2016,Smith2017}).  Masses between
several tens to a few hundred \(M_\odot\) could be populated by the
remnants of metal-free Population III stars, which are expected in
this mass range and are believed to have their peak formation rates
between \(z\sim5-8\) \cite{Tornatore2007,Johnson2012}. Observations
have revealed Pop III stars at \(z\sim6.5\) \cite{Sobral2015},
providing support for this picture. For masses \(25\)-\(140 M_\odot\)
and greater than \(260 M_\odot\) Pop III stars are expected to end
their lives as collapsars \cite{Heger2002,Heger2003}, producing failed supernova and BH-disk remnants suspected of powering distant, long gamma-ray
bursts~\cite{MacFadyen1999,MacFadyen2001}.  The least massive
potential progenitors are binary neutron star (NSNS) and neutron
star-BH (BHNS) mergers, the final BH masses of which are expected to
cover the approximate range \(3\)-\(20M_\odot\) \cite{Voss2003}.  We
point out that while an EOS with \(\Gamma=4/3\) is likely appropriate for
supermassive and Pop III stars
\cite{Shapiro2004,Shibata2016,Shibata2016a}, it is not appropriate for
NSNS and BHNS systems where nuclear matter is at play. Thus, when
applying our results to low-mass systems they should only be viewed as
approximate.

\begin{figure}[!tb]
\includegraphics{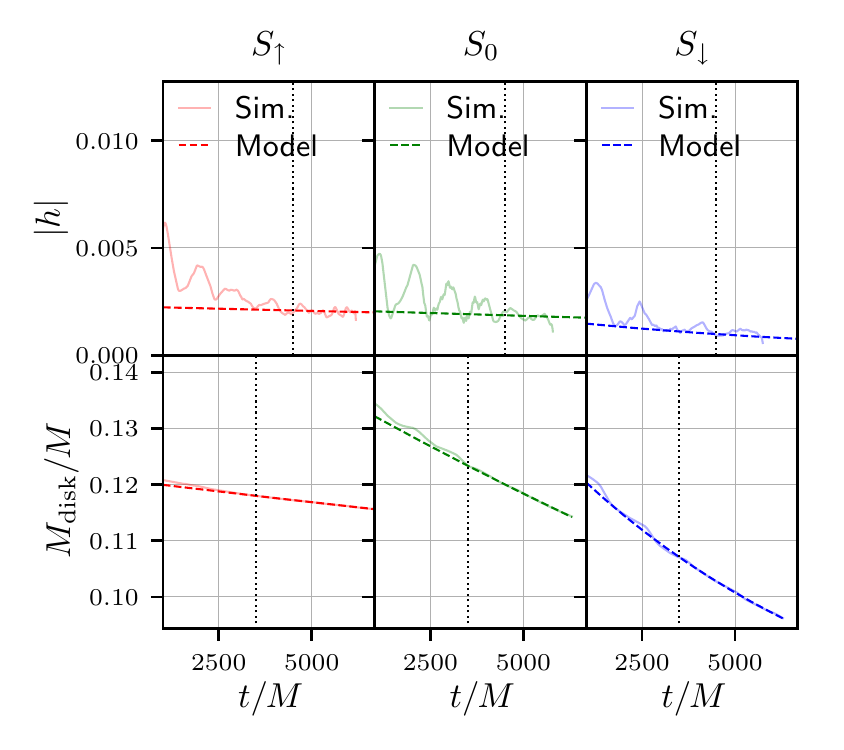}
	\caption{\label{fig:model_fitting_plots}Top row: Signal
          amplitude time evolution (faint solid lines) and model
          amplitude time evolution (dashed lines). Bottom row: Total
          disk mass (faint solid lines) and modeled disk mass (dashed
          lines). The model was fit to the data only in the regions
          to the right of the vertical dotted lines, to avoid
          interference by transient features in the GW
          signal and accretion rate.}
\end{figure}

After scaling to the appropriate mass scale, the signals are
propagated from the source frame to the observer frame through a flat
\(\Lambda\)CDM cosmology, and the characteristic strain is computed
(see equations \eqref{eq:char_str_def} and~\eqref{eq:pol_avg_def}). In
our analysis we remove the first \(\Delta t=1000 M\) of the signal to
eliminate the initial violent hydrodynamic relaxation of the
initial data as the instability develops. As in
Fig.~\ref{fig:comp_char_strain} we adopt an angle \(\theta =
\pi/2.34\) for the orbital inclination, which results in the dominant
\(l=2, m=2\) mode amplitude being equal to its $\theta$-averaged
value.  We then compute a ``sky-averaged'' signal-to-noise (SNR) for
such an event if observed by Advanced LIGO \cite{Collaboration2013},
Cosmic Explorer \cite{Reitze2019}, DECIGO \cite{Sato2017}, or LISA
\cite{AmaroSeoane2017}, assuming an optimal matched filter.
Sensitivity curves for the three ground-based observatories were
obtained from \cite{GroundNoiseCurves2016}, divided by the
sky-averaged antenna response function for a \(90^{\circ}\)
interferometer (see equation 51 of \cite{Moore2014}).  The analytic
approximations given in \cite{Yagi2011} and \cite{Robson2019} were
used for the DECIGO and LISA sky-averaged sensitivities,
respectively.\footnote{One technical complication arises: our
  definition of \(h_c\) already accounts for polarization averaging by
  dividing by a factor of \(\sqrt{2}\) in equation
  (\ref{eq:pol_avg_def}).  In order to keep the ratio of signal and
  sensitivity heights equal to the SNR (see caption of Figure
  \ref{fig:detection_horizon}), we multiply the sky \& polarization
  averaged sensitivities given by \cite{Moore2014,Yagi2011,Robson2019}
  by this factor of \(\sqrt{2}\) before plotting them, so that in
  effect the plotted sensitivities account only for the sky-position
  averaging, but not the polarization averaging, which is already
  included in \(h_c\).}

\begin{figure*}[!tb]
\includegraphics{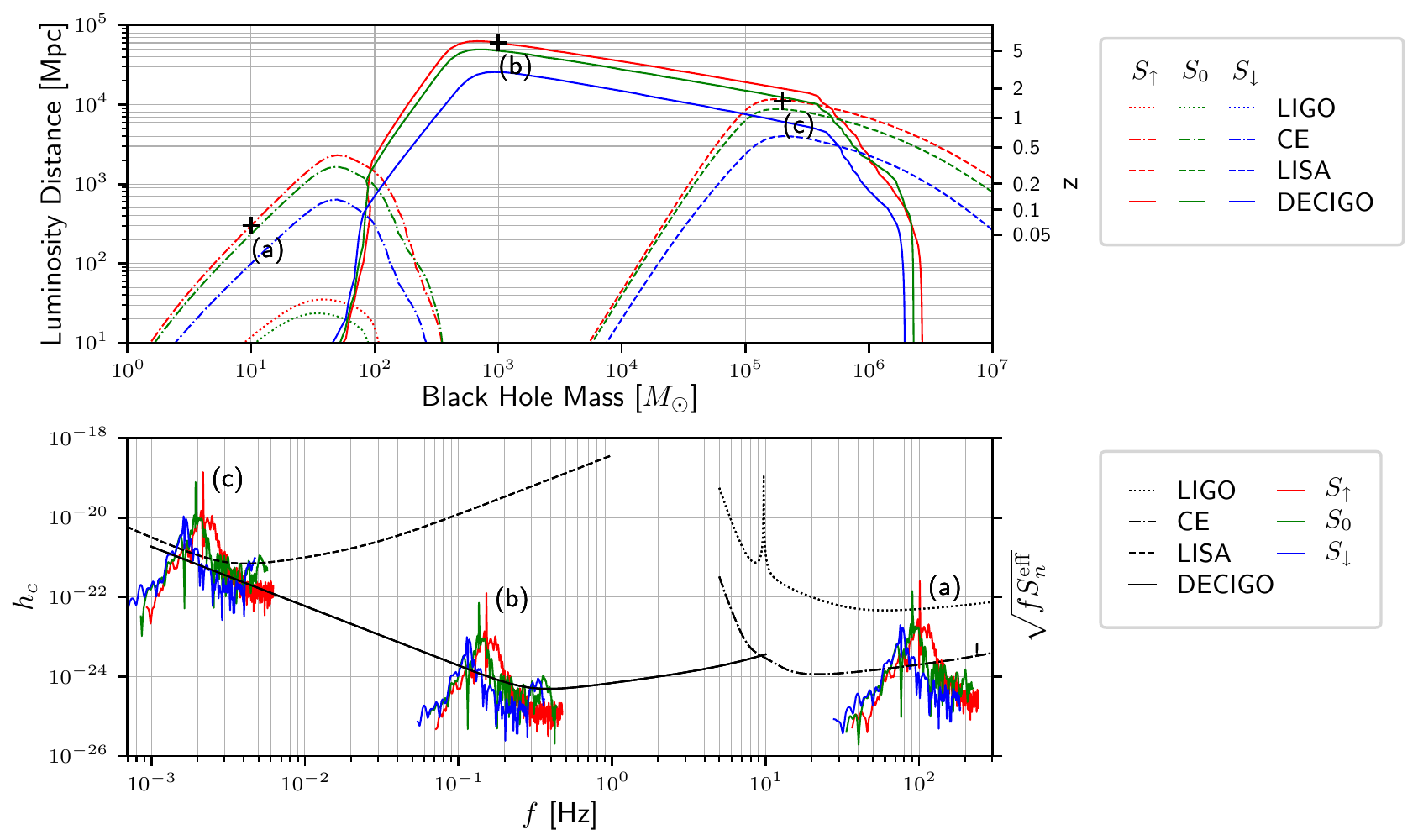}
\caption{\label{fig:detection_horizon_ext} Same as
  Fig.~\ref{fig:detection_horizon} but for the extrapolated signal.
  The three hypothetical sources representing the most distant systems
  detectable for the same three BH masses from Figure
  \ref{fig:detection_horizon} now become: (a) \(10 M_\odot\), \(300\)
  Mpc; (b) \(1000 M_\odot\), 60000 Mpc (\(z=6.08\)); (c) \(2\times10^5
  M_\odot\), \(11000\) Mpc (\(z=1.47\)).}
\end{figure*}

The results of the SNR calculation using the simulated part of the
GW signal (after \(1000 M\)) for each case are shown in
Figure~\ref{fig:detection_horizon}. The top panel shows the maximum
distance or redshift a system of given mass would be detectable
assuming an SNR detection threshold of 8. From the plot it becomes
clear that Advanced LIGO can detect such systems at a maximum distance
of just under \(\sim 20\) Mpc (for the case of a \(20 M_\odot\) BH). On the other
hand, Cosmic Explorer will be able to detect a \(20 M_\odot\) source
out to \(\sim 400\) Mpc, and can detect a 10 \(M_\odot\) system
(marked by a $+$ and labeled (a)) out to 150 Mpc. It is therefore
possible that future ground-based detectors could observe such
systems.

Space-based observatories will be able to detect more distant and
massive sources. As shown in Figure~\ref{fig:detection_horizon},
DECIGO and LISA are well suited to detect systems with masses
\(10^3\)-\(10^6 M_\odot\), and can detect them out to many Gpc. DECIGO
in particular, owing to its superb sensitivity, would be able to
detect such systems out to several tens of Gpc. A system with mass
$10^3M_\odot$ (labeled as source (b)) can be detected much further out
than any other system mass, with the maximum distance corresponding to
a cosmological redshift of $z=4.3$.  On the other hand, LISA will be
able to detect BH-disk systems with mass $10^5-10^6M_\odot$ out to
cosmological redshift of $z=1$ (for the source mass labeled (c)).  The
characteristic strain for sources with masses corresponding to those
labeled (a), (b), (c) in the top panel are shown in the bottom panel
for each of the three spin states we simulated. We also plot the
corresponding sensitivity curves of LIGO, Cosmic Explorer (labeled CE in the figure), DECIGO
and LISA.

In Figure~\ref{fig:detection_horizon}, we considered just the GW
signal that was extracted from in our simulations (excluding the
first \(1000 M\)).  However, at the end of our
simulations the disks still emit significant gravitational radiation,
and the orbiting \(m=1\) over-densities responsible for that radiation
appear to be stable features in all three cases. Consequently, we
expect that the disks will continue emitting a strong GW signal until
a significant amount of the rest-mass has been accreted, increasing
the total signal duration.

To obtain a better estimate of the detectability of BH-disk systems,
we therefore need to extrapolate beyond the portion of the signal that
was simulated. For simplicity, we restrict ourselves to modeling the
dominant \(l=2\), \(m=2\) mode. Motivated by Equation
\ref{eq:quad_signal}, we assume that the signal will be similar to
that of two orbiting point masses. Due to accretion, mass is slowly
transferred from the disk to the BH. The simulation data show that the
disk rest-mass decays approximately exponentially. This can be modeled
in the quadrupole formula by inserting \(\mu \propto e^{-\gamma t}\)
into Eq.~\ref{eq:quad_signal}. As long as \(\gamma\) is small relative
to the orbital frequency \(\Omega_c\), and we can ignore the time
derivatives due to mass transfer when taking time derivatives of the
quadrupole moment. Under these assumptions, the signal after the
transient period will be of the form

\begin{equation}\label{eq:sig_model}
h = Be^{(i\omega_0 - \gamma)t}.
\end{equation}

This form indeed matches the observed late-time behavior of our
simulated GW signal. To match this model to the observed waveforms we
first chose the value of \(\omega_0\) via a least-squares linear fit
to the unrolled phase of the complex \(l=2\), \(m=2\) strain. Since
the normalized \(m=1\) density mode amplitude does not appear to decay
over the duration of simulation, we assume that the dwindling mass of
the disk determines the signal amplitude falloff at late times. The
late-time decay rate parameter \(\gamma\) was extracted from the disk
mass evolution, rather than from the signal itself, by fitting the
late-time profile of the total disk mass. Finally, the amplitude,
\(B\), was chosen by a least-squares fit of \(B e^{-\gamma t}\) to the
late-time amplitude profile of the GW signal (with \(\gamma\) fixed to
the value found in the previous step). Figure
\ref{fig:model_fitting_plots} shows the fits.

To explore the limits of potential detectability, we assume that the
signal will persist until 90\% of the disk mass has been accreted,
after which the amplitude smoothly drops to zero over a few
orbits\footnote{Ending the GW signal after 50\% of the disk was
  accreted reduced the maximum detectable distance by a factor of
  \(\sim0.9\) compared to the 90\% case, so detectability is not
  sensitive to the termination threshold, because most of the SNR
  comes from the early part of the signal.}.

The detection horizon and characteristic strain of this extended
signal are shown in Figure~\ref{fig:detection_horizon_ext}.  By
extending the signal duration to a significant fraction of the
lifetime of the disk we raise the maximum detectable luminosity
distances for all three spin states, with \Su{} receiving the biggest
boost due to its long accretion timescale. Cosmic Explorer is now able
to detect the \(10 M_\odot\) \Su{} signal out to 300 Mpc, and a \(20
M_\odot\) source out to \(\sim 500\) Mpc. DECIGO and LISA can detect
sources in their frequency ranges 30\%-50\% further away, with the
most distant source (b) now detectable by DECIGO out to a redshift of
6.08.

Taken together, Figures \ref{fig:detection_horizon} and
\ref{fig:detection_horizon_ext} provide a range of the potential
detectability of PPI-unstable disks. We can therefore conclude that
PPI unstable disks similar to \Su{} are detectable by DECIGO out to
\(z \approx 5\), and around \(z \approx 1\) by LISA. For lower mass
systems, Cosmic Explorer will likely be able detect them out to
several hundred Mpc, with the limit of detectability of a system with
a \(10 M_\odot\) BH around \(\sim 300\) Mpc, which is near the
estimated distance of two confirmed LIGO binary merger
detections~\cite{Abbott2017,Abbott2020}, making it a realistic distance
to expect black hole-neutron star mergers.

\section{Discussion\label{section:discuss}}

Instabilities in BH accretion disks can result in time-changing
quadrupole moments and hence result in copious emission of GWs. We
embarked on a comprehensive study of such events, starting with the
PPI as a promising multi-messenger candidate for future ground-based
and space-based GW observatories. We consider the PPI in BH-disk
systems where the BH is spinning, and perform hydrodynamic simulations
in full general relativity starting with equilibrium and constraint
satisfying initial data. When the BH spin is aligned (case \Su{}) or
anti-aligned (case \Sd{}) with the disk's orbital angular momentum,
our simulations demonstrate the dynamics of PPI growth and saturation
does not differ significantly from the previously studied non-spinning
case (labeled \So{}). All three disks grew \(m=1\) instabilities on
similar timescales, and saturated to a similar \(m=1\) state. The
dominant frequencies in the non-axisymmetric density mode spectra were
proportional to each disk's orbital frequency at maximum density
(\(\Omega_c\)), and the spectra align almost perfectly once this
frequency re-scaling was accounted for. This was also true for the GW
signal, except that the spin-aligned case also had slightly higher
amplitudes than the non-spinning case, while the spin anti-aligned
case had slightly lower amplitudes. This behavior is consistent with
expectations from the quadrupole formula of orbiting masses, where the
amplitude is proportional to the square of the orbital velocity (see
Equation \ref{eq:quad_signal}).

Due to hydrodynamic shocks arising as the instability grows and
saturates, violent re-arrangement of the disk profile takes place
which leads to angular momentum re-distribution that allows accretion
to proceed. Assuming that 1\% of the accretion power in the relaxed
state is converted to bolometric electromagnetic luminosity, we
estimate electromagnetic counterparts as bright as
\(\mathcal{O}(10^{52})\)[erg/s]. Such high luminosities would be
detectable at very long distances assuming the conversion efficiency
to observable electromagnetic frequencies is not very small. In case
\Su{}, where the BH spin was aligned with the disk orbital angular
momentum, we saw a significant reduction of the accretion, with rates
over an order of magnitude suppressed relative to \So{} and \Sd{}, and
significantly lower than those previously reported for non-spinning
black holes \cite{Kiuchi11}.  This effect correlates well with
differences in distance between the inner edges of the disks and the
radii of the innermost stable circular orbit: the initial inner edge
of the disk in \Su{} is much farther from the ISCO than the disks in
\So{} and \Sd{}.  An exploration of various initial accretion disk
profiles would need to be undertaken to determine how and whether the
accretion rate for a PPI unstable disk can be used to measure BH spin.

While the PPI itself appears unaffected by BH spin, spin has indirect
impact on the frequency of the dominant PPI modes by affecting the
orbital frequency at maximum rest-mass density, and can significantly
impact the accretion rate. Thus, BH spin can act as a new degree of
freedom for controlling the lifetime of a disk undergoing the
PPI. This can significantly increase the lifetime of GW signals
emanating from PPI-unstable BH-disk systems, thus increasing their
detectability. However, the effects of magnetic fields should be
considered for a reliable measurement of the accretion rate, and the
subsequent lifetime of the PPI unstable mode. This will be the topic
of future work.

We applied our simulation results to a range of masses, focusing
primarily on two categories of potential BH-disk systems: compact
binary remnants and supermassive collapsing stars. While not
detectable by Advanced LIGO, the larger scale black hole-neutron star
merger remnants are promising candidates for detection by Cosmic
Explorer, which could detect GWs from a \(\approx 1 M_\odot\) PPI
unstable disk around a 10 \(M_\odot\) BH out to 150-300 Mpc. The
proposed space-based DECIGO mission seems to be ideally positioned to
detect supermassive star remnants massing \(\mathcal{O}(10^3)
M_\odot\), which it can detect out to redshift of \(z\sim5\). While
LISA could also detect the supermassive star with mass
\(\mathcal{O}(10^{5-6}) M_\odot\), it lacks the sensitivity to detect
them at redshift much larger than \(z=1\).

Our work demonstrates that disk instabilities can be promising sources
for coincident electromagnetic and GW detections by future GW
observatories. The near quasi-monochromatic GWs from PPI unstable
systems will make it straightforward to design templates for
detection. In a forthcoming paper we will present the results from
dynamical spacetime hydrodynamic simulations of misaligned BH-disk
systems.

\begin{acknowledgments}
This work was supported by NSF Grant PHY-1912619 to the University of
Arizona, and by NSF Grants No. PHY-1662211 and No. PHY-2006066, and
NASA Grant No. 80NSSC17K0070 to the University of Illinois at
Urbana-Champaign. High performance computing (HPC) resources were
provided by the Extreme Science and Engineering Discovery Environment
(XSEDE) under grant number TG-PHY190020. XSEDE is supported by NSF
grant No. ACI-1548562. Simulations and data analyses were performed
with the following resources: Stampede2 cluster provided by the Texas
Advanced Computing Center (TACC) at The University of Texas at Austin,
which is funded by the NSF through award ACI-1540931, and the the
Ocelote cluster at the University of Arizona, supported by the
UArizona TRIF, UITS, and Research, Innovation, and Impact (RII) and
maintained by the UArizona Research Technologies department.
\end{acknowledgments}

\bibliography{PPIBibliography}

\end{document}